# Fast and efficient photodetection in nanoscale quantum-dot junctions

*Ferry Prins[1*†], Michele Buscema[1], Johannes S. Seldenthuis[1], Samir Etaki[1], Gilles Buchs[1], Maria Barkelid[1], Val Zwiller[1], Yunan Gao[1,2], Arjan J. Houtepen[2], Laurens D. A. Siebbeles[2], Herre S. J. van der Zant[1*]*

1 Kavli Institute of Nanoscience, Delft University of Technology, PO Box 5046, 2600 GA, Delft, The Netherlands.

2 Department of Chemical Engineering, Delft University of Technology, Julianalaan 136, 2628 BL Delft, The Netherlands.

ABSTRACT We report on a photodetector in which colloidal quantum-dots directly bridge nanometer-spaced electrodes. Unlike in conventional quantum-dot thin film photodetectors, charge mobility no longer plays a role in our quantum-dot junctions as charge extraction requires only two individual tunnel events. We find an efficient photoconductive gain mechanism with external quantum-efficiencies of 38 electrons-per-photon in combination with response times faster than 300 ns. This compact device-architecture may open up new routes for improved photodetector performance in which efficiency and bandwidth do not go at the cost of one another.

Colloidal quantum dots (CQDs) profit from the quantum size effect[1], which gives rise to a variety of unique phenomena such as size-tunability, multi-exciton processes[2-4] and slow carrier-relaxation[5,6]. Moreover, they combine low-temperature synthetic methodology with solution processability, allowing for low-cost fabrication methods[7,8]. To date, the use of CQDs as photosensitive material has been mainly focused on thin-film devices[4,7-9]. In these systems, the extraction-efficiencies of photogenerated charges are dominated by the charge mobility, which in CQD-films is generally described by hopping through interparticle barriers. Techniques to enhance the electronic coupling between the CQDs and thus improve the film-mobility include the use of short linkers[10], various surface passivation approaches[11,12] and the use of CQDs with large Bohr-radii, in particular lead chalcogenide quantum dots. In addition, short channel-length detectors have shown promising performances[13,14].

An important advancement in the field of quantum-dot photodetectors has been the development of devices capable of photoconductive gain [7,9]. In these devices, exciton generation is followed by the trapping of one of the charge carriers, thereby lowering the chemical potential for transport of the opposite charge carrier. If the trap life-time exceeds the transit time of the opposite charge carrier, many carriers worth of current can pass through the circuit before recombination takes place. As a consequence, the measured photoconductance is considerably larger than without the presence of such a gain mechanism. This gain in efficiency does, however, go at the expense of an increase in the response time due to the slow trap-state dynamics, which is typically in the order of milliseconds[7,9].

Here, we present a photodetector which places a one-dimensional parallel array of quantum dots in direct contact with nanometer separated electrodes (see Fig. 1a). In this CQD-junction, charge mobility no longer plays a role as the contact of both source and drain electrodes to each CQD allows for direct charge extraction, which is both fast and efficient.

Nanometer-spaced electrodes are fabricated by a self-aligned fabrication scheme, consisting of a basic two-step lithography process[15-17] (see supporting information for details). An advantage of the self-alignment technique is that the nanometer-separated electrodes can be prepared over large widths, which allows contact to many particles in parallel (see Fig. 1b). Here, we use devices with an electrode separation of 4 nm, and with an electrode width of 10 μm. A single layer (see supplementary information) of PbSe quantum dots of 4 nm in size is placed on top of the electrodes using dipcoating[18] after which a subsequent ligand substitution step with 1,2-ethanedithiol increases the coupling to the electrodes. Electrical contact of the particles to the electrodes is confirmed by performing electrical characterization before and after deposition at room temperature in a vacuum probe station. Before deposition the resistance in all 300 devices studied is > 100 GΩ at voltages up to 2.5 V, whereas after deposition of the PbSe QDs a clear onset of conductance is observed at approximately 1 V, reflecting the density of states inside the dots (see Fig. 1c).

In all devices, we found a strong photoconductive effect when irradiating them with visible light. Current-voltage (I-V) characteristics taken under laser light irradiation (λ = 532 nm, and an irradiance of E = 0.16 Wcm$^{-2}$, see Fig. 2a) display a linear dependence with the applied voltage in the low-bias regime; at higher bias the I-V characteristics become non-linear at the onset of the dark current. Control experiments in which bare nanogaps without PbSe-QDs are illuminated, display no photoconductive response.

To spatially resolve the photoconductive response, we place a device in an optical scanning confocal-microscope setup. While scanning a diffraction limited laser spot ($\lambda$ = 532 nm, spot size of ~800 nm) across the device, the current is measured as a function of the spot position. Conductance maps, recorded at 750 mV, show a high photoconductive response when the laser spot is placed directly on top of the nanogap area (see Fig. 2b, data corrected for the dark current). The response along the gap is consistently above 2 nA, with some variations in the current (within a factor of two). These variations maybe the result of a non-uniformity in the distribution of PbSe QDs along the gap or in the electronic coupling of the particles to the electrodes. The full width at half maximum of the photoresponse is 805 nm measured perpendicular to the nanogap, consistent with the spot-size of our diffraction-limited laser beam (see supporting information).

The photoconductive response depends strongly on the wavelength of the incident light. For the device of Fig. 2b, we fix the laser at the position of maximum response and measure the current at V = 750 mV and constant optical power, P = 15 μW, for varying wavelengths between 850 and 1650 nm as shown in Fig. 2c. The spectral dependence of the photoconductance closely resembles the absorption spectrum of a reference film of the same PbSe QDs on quartz (grey solid line in Fig. 2c), including the peak absorption at the bandgap energy and the absence of response at longer wavelengths ($\lambda$ > 1600 nm). The correspondence to the QD's absorption spectrum shows that the photoconductance is driven by optical excitations inside the QDs.

We have also recorded the photocurrent at $\lambda$ = 532 nm at different optical powers (between E = 2 and 2000 Wcm$^{-2}$, shown in Fig. 3a). At low powers (E < 50 Wcm$^{-2}$), the current scales linearly with the laser power, while at higher powers (E > 50 Wcm$^{-2}$) the current saturates. We attribute the saturation at high powers to the filling of trap states which reduces the efficiency to the photoconductance gain mechanism, as is discussed below. At lower powers the efficiency is highest. The external quantum efficiency (EQE), i.e. the number of extracted electrons (or holes) per photon incident on the device, can be calculated with

$$EQE = \frac{I_{ph}}{e} \frac{\hbar\omega}{E \cdot A} , \qquad (1)$$

where $I_{ph}$ is photocurrent (corrected for the dark current) and A is the illuminated device area (4 x 800 nm$^2$). For the device in Fig 3a, the efficiency reaches 10.9 electrons per photon at 2 Wcm$^{-2}$ with a bias of 750 mV (see supporting information). At higher bias the photon-to-electron conversion is even more efficient; for example in the I-V characteristic of Fig. 2a, which is taken at low power, an EQE of 38 electrons per photon is observed at a bias of 1.5 V. In the calculation of the EQE it is assumed that all light incident on the gap area is absorbed by the one-dimensional array of quantum dots. If we do take the absorption probability of the single row of CQDs into account[19], we can determine the internal quantum efficiency (IQE) of the detector. Assuming a coverage of 2000 4-nm-sized particles across the 10 μm wide device of Fig. 2a, we obtain IQEs as high as 5.9 x 10$^3$ electrons per photon (see supporting information).

The most striking feature of our device performance is the combination of high efficiency with fast response. Fig. 3b shows the temporal photoresponse of the device to a 1 μs laser pulse, measured using a low-noise amplifier and a 600 MHz oscilloscope. The observed rise and fall time of our device are approximately 200 and 300 ns, respectively, limited by the electronic bandwidth of the measurement circuit (see supporting information). Such fast response times combined with high efficiencies are crucial parameters in, for instance, video-rate laser-scanning microscopy where dwell-times in the order of 100 ns are required[20].

High efficiencies in CQD-photodetectors are usually achieved by trapping of one of the charge carriers, which is photoinduced by the incident light. The trapping of charge carrier lowers the chemical potential for transport of the opposite charge carrier, thereby increasing the conductance[8]. If the lifetime of the trap state exceeds the carrier transit-time, several opposite charge carriers can pass through the device before recombination with the trapped charge occurs. Thus, instead of the charge carriers that originate from the exciton pair only, many more charge carriers contribute to the photoconductance, The gain factor in this picture is proportional to the ratio of the trap lifetime over the transit time. For conventional film-based CQD photoconductors, which have typical mobilities between $10^{-1}$ and $10^{-3}$ $cm^2Vs^{-1}$ and channel lengths of around a micron, the transit time is hundreds of nanoseconds or longer[12,13]. Appreciable gain therefore requires trap-state life times as long as milliseconds, inevitably leading to the slow response times of these devices.

The orders-of-magnitude shorter rise and fall time of our device shows that if trap-induced photoconductive gain is responsible for the high efficiency, the mechanism needs to be active on very short time scales. This would first of all require short-lived trap states, shorter than the observed 200 and 300 ns rise and fall time, and second, it would require sub-nanosecond transit times in order to explain the observed IQE. Short transit times are indeed feasible in our device as a direct result of the nanoscale geometry: the transit time is determined by the product of the tunnelling probabilities of the two barriers that separate each CQD from source and drain. The interfacial electronic coupling energy for PbSe CQDs at or near resonance is reported to be as high as 100 meV[21], corresponding to individual tunnel events at sub-picosecond timescales[6]. Taking this into consideration, even short lived traps (<< 200 ns) could thus lead to considerable gain factors in our device, allowing for high efficiency at short timescales.

For a possible explanation for the short-lived trap states we take a closer look at the quantum dot junction. With the CQD only separated from source and drain by two tunnel barriers, the stochastic nature of the individual charge extraction events becomes important[22-24]. After creation of an exciton, either the electron or the hole will be extracted first, leaving the opposite carrier temporarily behind as if it was trapped. If it is for instance the hole which is left behind, the chemical potential for transport of electrons through the dot is brought closer into resonance with the Fermi level as a result of the reduced Coulombic repulsion. This leads to photoconductive gain in the same way as a conventional trap state at for instance the surface of a quantum dot would.[25] The crucial difference with conventional traps is, however, that the lifetime of the trap is only determined by the probabilistic nature of the tunnel events, which occur at the before mentioned sub-picosecond time-scales[6, 21].

In the above discussion we have not taken into account plasmonic field enhancement effects that are known to increase light collection in nanogapped structures (i.e., enlarge the optical cross-section of the dots)[26-30]. If present, they would also contribute to the enhancement of the IQE number so that the gain is not solely determined by the ratio of the trap lifetime over the transit time.

In conclusion, the presented architecture offers a promising route towards solution processable, low-cost, nanoscale devices with ultrafast yet efficient detection performance. It moreover comprises a versatile platform to study the microscopic details of charge transfer at the metallic interface, not limited to colloidal quantum dots only, but applicable to a variety of nanomaterials. In addition, we expect that plasmonic field enhancement may already play a role in the efficiency of our device, although the effect is expected to be small for the broad non-resonant electrodes we have used in our experiment. It will therefore be interesting to investigate devices with smaller widths which could profit from both efficient carrier extraction, as well as optimized light absorption by resonant plasmonic field enhancement.

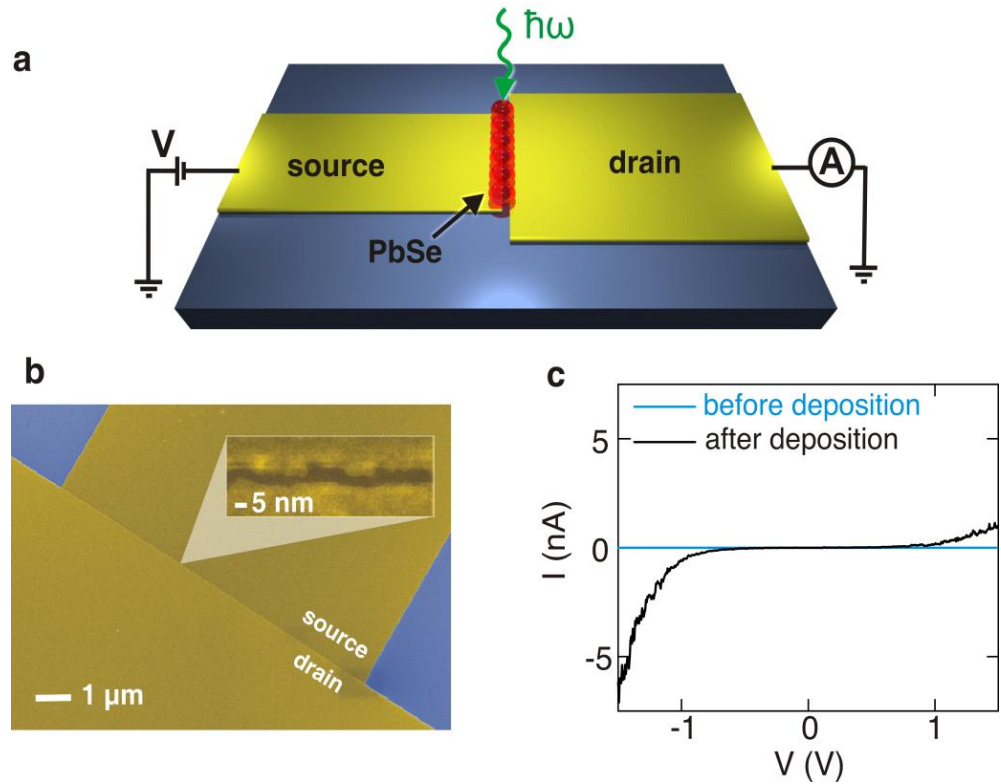

**Figure 1.** (a) Schematic of the device architecture illustrating the one-dimensional CQD array geometry. (b) Colorized Scanning Electron Micrograph of an empty 10 μm wide device fabricated by a self-aligned fabrication scheme (see methods section). The inset shows a zoom-in of the nanogap area. (c) Room-temperature current-voltage characteristics before and after deposition of PbSe CQDs. The device width is 10 μm. Measurements are performed in a vacuum probe-station at room temperature.

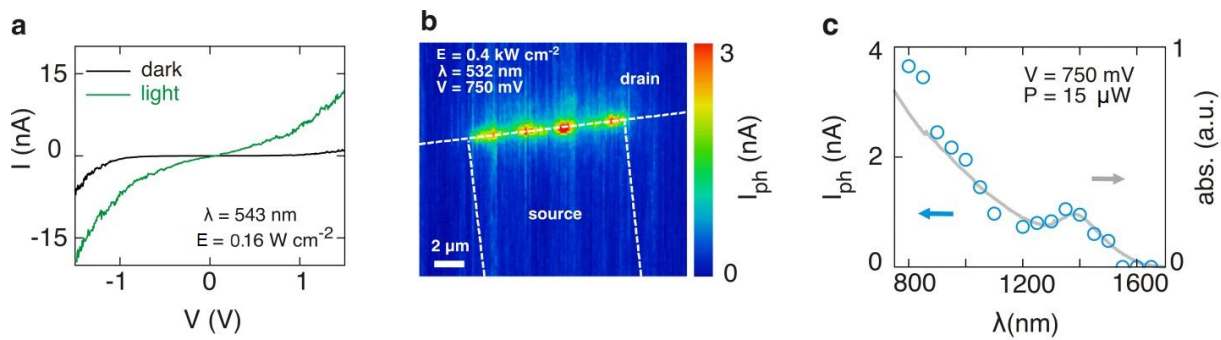

**Figure 2.** (a) I-V characteristics in the dark (solid black line) and under laser illumination (solid green line, λ = 543 nm, E = 0.16 Wcm$^{-2}$, spot size = 150 μm.) (b) Map of the photocurrent of a different device (corrected for a dark current of 0.8 nA) as a function of the position of the diffraction-limited spot from a different setup with a spot size of about 800 nm (λ = 532 nm). Dashed white lines indicate the electrode edges determined from the reflection image. (c) Wavelength dependence of the photocurrent of the device in B at constant laser power (blue open circles). Photocurrent points are taken at a fixed position of maximum response of the device. The optical absorption spectrum of a solid film of CQDs (solid grey line) is shown for comparison.

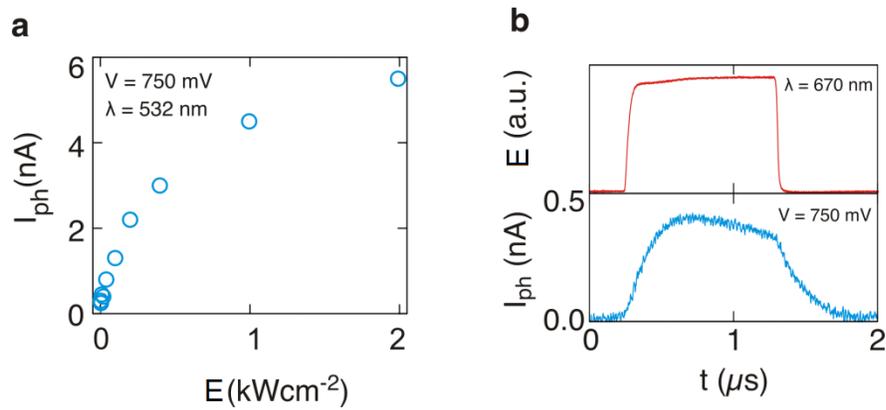

**Figure 3.** (a) photocurrent as a function of irradiance (blue open circles, corrected for the dark current) for the device in Fig. 2b and c. (b) photocurrent response (bottom panel, corrected for the dark current) of a different device to a 1 μs square pulsed laser illumination (top panel, λ = 670 nm, E = 0.85 Wcm$^{-2}$, spot size = 150 μm). The rise and fall time of the laser signal is < 20 ns.

**Supporting Information**. Experimental details, additional device characterization, bandwidth analysis.

**Corresponding Author**

*E-Mail: (F.P.) prins@mit.edu; (H.vd.Z.) h.s.j.vanderzant@tudelft.nl

**Present Addresses**

† Department of Chemical Engineering, Massachusetts Institute of Technology, 77 Massachusetts Avenue, Cambridge, Massachusetts 02139, United States.

ACKNOWLEDGMENT

We thank G. A. Steele, J. M. Thijssen, Y. M. Blanter, R. W. Heeres and W. A. Tisdale for discussions. We thank M. van Oossanen, R. Schouten and R. van Ooijik for technical assistance. This work was supported by Stichting FOM under the programs 86 and 111 and a FOM projectruimte, through the FP7-framework program ELFOS and a Marie Curie Intra European Fellowship (G.B.).

# Supporting information

# Fast and efficient photodetection in nanoscale quantum-dot junctions

*Ferry Prins[1*†], Michele Buscema[1], Johannes S. Seldenthuis[1], Samir Etaki[1], Gilles Buchs[1], Maria Barkelid[1], Val Zwiller[1], Yunan Gao[1,2], Arjan J. Houtepen[2], Laurens D. A. Siebbeles[2], Herre S. J. van der Zant[1*]*

1 Kavli Institute of Nanoscience, Delft University of Technology, PO Box 5046, 2600 GA, Delft, The Netherlands.

2 Department of Chemical Engineering, Delft University of Technology, Julianalaan 136, 2628 BL Delft, The Netherlands.

DEVICE FABRICATION- Electrodes were fabricated in three steps by standard electron-beam lithography techniques. On top of a Si/SiO$_2$ substrate, the first electrode is defined and deposited by electron-beam evaporation, consisting of a 2 nm thick Ti adhesive layer, with 24 nm Au on top as the electrode material, and 10 nm of Cr. Upon removal from the evaporation chamber, the Cr forms a natural Cr$_2$O$_3$ layer which overhangs the Au as a shadow mask. In the second step, the second electrode of 2 nm Ti with 20 nm Au on top is defined, overlaying the first Ti/Au/Cr/Cr$_2$O$_3$ electrode, while the shadow mask protects a few nanometers around the first electrode. In the third step, a selective wet Cr-etchant (Cyantek, Cr-10, 5 minutes immersion) is used to remove the shadowmask, exposing a nanometer sized separation between the first and second electrode (3 - 6 nm as determined from scanning electron microscopy, see inset Fig. 1b). Before use, the samples are O$_2$-plasma cleaned.

CQD SYNTHESIS AND DEPOSITION- PbSe CQDs of appr. 4 nm in diameter were synthesized according to ref.1. PbSe particles are deposited on the electrodes by a dipcoating technique under N$_2$-atmosphere. First, the devices are immersed in a MeOH solution of ethylenedithiol for 3 hours to mediate a chemical linkage between the Au and the PbSe particles. The devices are subsequently immersed in MeOH (1 min), a hexane solution of oleylamine-coated PbSe nanoparticles (1 min). Finally, the devices are once more immersed in a MeOH solution of ethylenedithiol (1 min, followed by 1 min in pure MeOH), to substitute the long oleylamine ligands and minimize the barrier between the particles and the electrodes. Figure S1 shows an AFM micrograph of the resulting PbSe film over a gold electrode in a self-aligned device. The film is uniform over more than 10 μm, showing only small rim-like structures of higher elevation. The line trace in panel b) shows the step profile between the film and an area where the probes (used to make electrical contact to the pad) have removed the film itself. The step height is in the order of 5 nm, indicating that the film is composed of a single monolayer of PbSe QDs.

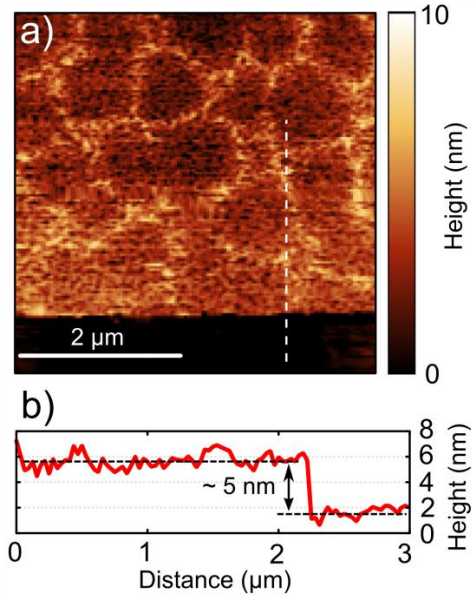

**Fig. S1. a)** AFM image of the PbSe CQD film deposited by layer-by-layer deposition (capped with EDT) on the surface of one of the gold electrodes in a self-aligned device. The region were the film is absent is due to contact with the electrical probes. The image was taken with a Nanoscope IIIa AFM in amplitude modulation mode with an Olympus OMCL-AC160TS-R3 cantilever. Scale bar is 2 μm. **b)** Line profile along the dotted line in panel (a).

DEVICE CHARACTERIZATION- Basic characterization was performed in a vacuum ($10^{-5}$ mbar) probe station (Desert cryogenics). Electrical measurements were performed using home-built electronics. Time trace measurements were performed using low-noise amplifier and a 600 MHz oscilloscope (LeCroy). Photoconductance measurements were performed by shining a CW-laser ($\lambda$ = 543 nm, 150 μm spot size) through the window of the probe station. Pulsed laser signals were generated by applying a microsecond block wave to a photodiode ($\lambda$ = 670 nm). The pulse shape was measured with a photodiode on a 600 MHz oscilloscope. Photoconductance maps were recorded in a vacuum station ($10^{-5}$ mbar) on a wire-bonded sample using a confocal microscope with a NA = 0.8 objective illuminated by a $\lambda$ = 532 nm CW-laser beam or by a filtered supercontinuum laser (Fianium, $\lambda$ = 800 – 1650 nm range). The diffraction limited spot is scanned using a combination of two galvo-mirrors and a telecentric lens system while the dc photoconductance signal and the reflected light intensity are recorded simultaneously in order to determine the absolute position of the detected photoconductance features. The optical powers were measured using Si and Ge photodiode power sensors (Thorlabs, S120C and S122C) placed before the final objective.

Before deposition, all devices have resistances > 100 GΩ at voltages up to 2.5 V, whereas after deposition of the PbSe QDs a clear onset in conductance is observed at approximately 1 V. An important advantage of the self-aligned fabrication technique is the possibility to prepare electrode-pairs with nanometer separation over large widths, i.e. creating trenches. In order to characterize the fabrication procedure further, we prepared devices with widths between 100 nm and 10 µm, in order to vary the maximum number of CQDs contacted in parallel. When plotting the average current of different devices (50 in total, 3 different depositions) at a bias of 1 V as a function of the device width, a roughly linear scaling is observed illustrating the reproducibility of the deposition of the CQDs on the devices (see Fig. S2).

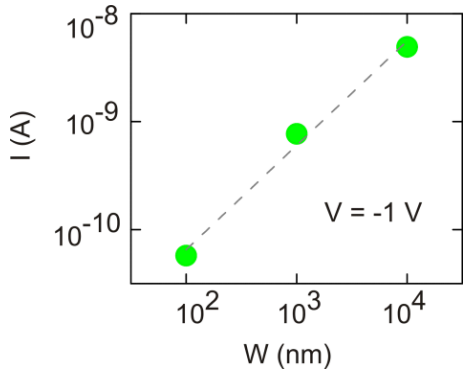

**Fig. S2** Average dark current at -1V as a function of the device width (i.e. number of particles in parallel).

Interestingly, we consistently observe asymmetric I-V characteristics with a lower current at positive bias (using the low electrode as the source). This is likely a result of the asymmetric height of the electrodes (see device fabrication) where the source is ~ 5 nm lower than the drain resulting in a difference in the electronic coupling.

Line traces taken perpendicular to the gap area in the conductance maps show the spatial distribution of the photoconductive response. The full width at half maximum of the conductance peak centered around the gap is 805 nm, consistent with the spot size of the diffraction limited spot (see Fig. S3).

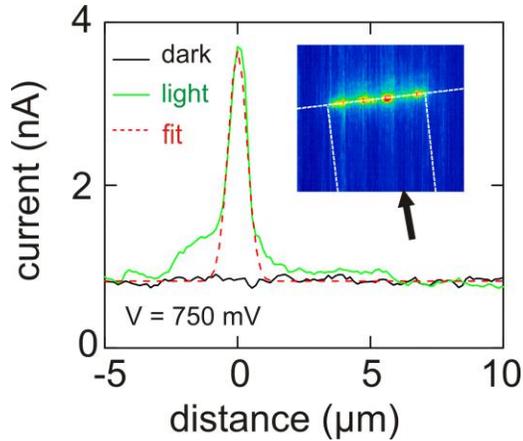

**Fig. S3** line-cut of the conductance map taken perpendicular to the nanogap (see arrow in inset). Same device as Fig 2b. The red dashed line is a Gaussian fit to the photocurrent response at the gap area. The full width at half maximum of the fit equals 805 nm, closely matching the spot size of the diffraction limited spot of the scanning confocal microscope at λ = 532 nm. Slight deviations to the Gaussian fit are observed at larger distances from the gap. This is likely the result of reflections from the underlying Si-surface since the thin Au electrodes are partially transparent.

QUANTUM EFFICIENCY CALCULATIONS- For the device in Fig 3a, the external quantum efficiency (EQE) is calculated according to equation 2 (see main text), taking the nanogap area within the spot size of 800 nm as the device area (800 x 4 = 3200 nm$^2$). At low power, the EQE reaches a maximum of 10.9 electrons per photon at 2 Wcm$^{-2}$, corresponding to a responsivity of 3.9 A/W (see Fig. S4). Similarly, the EQE for the device in Fig. 2c, in which the whole device is illuminated, can be calculated taking the full nanogap area into account (10,000 x 4 = 40,000 nm$^2$), obtaining an EQE of 38 electrons per photon with a power density of 1.6 Wcm$^{-2}$.

To determine the internal quantum efficiency (IQE) of the device we consider the optical cross-section of the sum of the maximum number of CQDs present in the gap area. For a 10 μm wide device, with a particle diameter of 4 nm, the maximum number of CQDs to be contacted in parallel is 2500. Taking the absorption cross-section of each particle to be 0.102 nm$^2$ at the excitation wavelength, as reported in ref. 2, we obtain a total optical cross section of a 10 μm wide device of 2500 x 0.102 = 255 nm$^2$. With a surface power density of 0.16 Wcm$^{-2}$ (4.5 x 10$^4$ $n_{ph}$ / s / nm$^2$) and a photocurrent of 10.9 nA at 1.5 V (68.1 x 10$^9$ $n_e$ / s, corrected for the dark current) we obtain an IQE of 5.9 x 10$^3$ for the device in Fig. 2a. It should be noted that the estimation of the number of particles is an upper limit and in reality is likely to be lower as a result of the distribution of the particles across the gap.

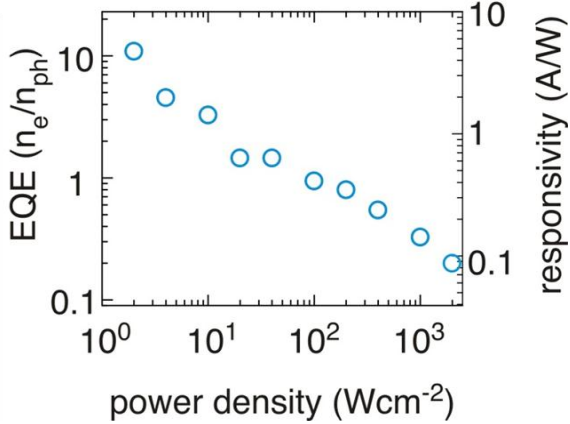

**Fig. S4** External quantum efficiency (EQE) and responsivity in terms of extracted electrons per photon incident on the illuminated device area (800 nm x 4 nm) of the device in Fig. 3a as a function of the surface power density.

BANDWIDTH DETERMINATION- Here we develop a model for the nanoparticle junction and the measurement setup. We characterize the post-amplifier and show that the measured rise time ($T_m$ = 189 ns from 10 % to 90 %) of the step response in figure 3b (see main text) is likely determined by the amplifier rise time ($T_h$) and not by that of the junction ($T_d$). Figure S5a shows the electronic circuit to be analysed. The junction, represented by a device resistance ($R_d$) and a capacitance to ground ($C_d$), is biased by a constant voltage source ($V_d$). The laser light (green waves) changes the device resistance. To determine $T_m$, the laser is pulsed and the voltage ($V_h$) is measured across the input resistor ($R_h$) of a low-noise HEMT (High Electron Mobility Transistor) amplifier. The capacitance from the device ($C_d$) and setup wiring ($C_c$) run parallel to $R_H$. The custom-built laser has a 5 ns rise time and the amplified voltage is measured using a fast oscilloscope (Rohde&Schwarz RTO 1014, 1 GHz, 10 GS/s). These elements are much faster than $T_m$ and are therefore not included in the schematic. The HEMT transistor has a high impedance compared to $R_h$ and is not considered here.

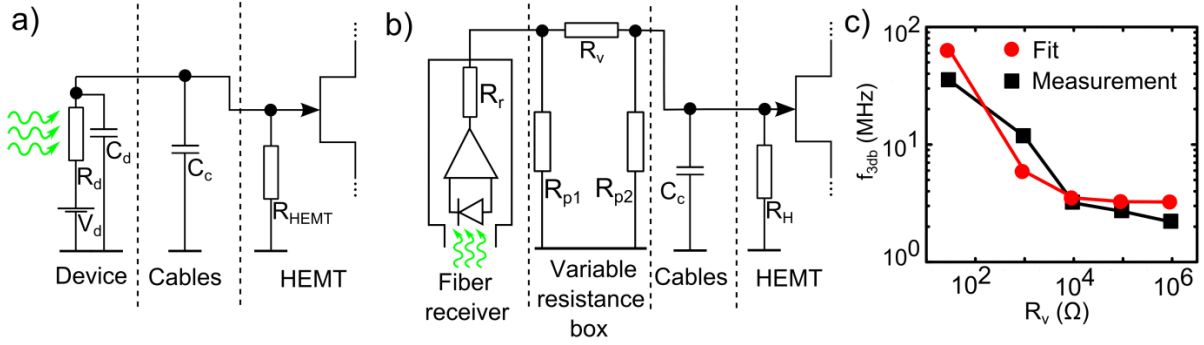

Fig. S5. a) Schematic representation of the measurement circuit, consisting of a nanoparticle junction device and a HEMT post-amplifier. The green waves represent the laser source which shines on the junction. b) Setup to determine the amplifier bandwidth. The junction is replaced by a fiber receiver and a variable series resistance embedded in a box. c) Measured (black squares) and theoretical (red circles) cut-off frequency of circuit in panel (b) as a function of the series resistance.

Analysis of the resulting circuit yields the following differential equation for $V_h$:

$$\{1\} \quad R_h(C_d + C_c)\dot{V}_h + \left(1 + \frac{R_h}{R_d}\right) V_h = \frac{R_h}{R_d} V_d \ .$$

The nanoparticle junction has $R_d$ = 100 MΩ under continuous wave illumination (see main text) and becomes an open connection in the dark. The amplifier has $R_h$ = 5 kΩ, so $R_h/R_d \ll 1$, which means that this term can be discarded on the left hand side of equation 1. Assuming $T_d \ll T_h$, the circuit is then equivalent to a first-order low-pass filter. The -3dB cutoff frequency is given by:

$$\{2\} \quad f_{3dB} = \frac{1}{2\pi R_h (C_d + C_c)} \ ,$$

And the 10 % to 90 % rise time is related to the cut-off frequency by:

$$\{3\} \quad T_h = \frac{0.35}{f_{3dB}} = 2.2 R_h (C_d + C_c) \ .$$

We determine $T_h$ from the device response to a step in the light amplitude. From equation 1 we see that, in general, the response of $V_h$ to an AC light signal is not the same as the response to an AC voltage $V_d$. However, the step response is the same for light and voltage (still assuming $T_d \ll T_h$) and we therefore use equations 2 and 3 for both. Although $R_h$ is known, $C_d$ and $C_c$ are not. Next, we perform a measurement to determine $C_c$. $C_d$ cannot be measured directly, but $C_c$ is expected to be dominant.

We determine $C_c$ by replacing the junction device with a commercial fiber receiver (figure S5b, AVAGO HFBR-2406Z, 5 ns rise time). The receiver acts as an AC voltage source with a low output resistance ($R_r$ = 30 Ω) and there is a variable series resistance ($R_v$) which represents the device. The resistances in parallel to ground ($R_{p1}$ = 30 Ω and $R_{p2}$ = 1 k Ω ) are used to preserve the pulse shape going into the setup wiring ($C_c$ in the schematics). The cut-off frequency now becomes:

$$\{4\} \quad f_{3dB} = \frac{1}{2\pi C_c} \frac{R_{p2} \cdot R_H + R_H \cdot (R_{eq} + R_v) + R_{p2} \cdot (R_{eq} + R_v)}{R_{p2} \cdot (R_{eq} + R_v) \cdot R_H}$$

where the symbols refer to figure S5b and $\frac{1}{R_{eq}} = \frac{1}{R_r} + \frac{1}{R_{p1}}$.

We determine $C_c$ by performing pulse measurements with several values of $R_v$, extracting the cutoff frequency from the rise time using equation 3 and fitting the result to equation 4. Figure S5c shows a good match between the measured and fitted values, which indicates that our model describes the measurement setup well. The fit yields a value $C_c$ = 26 pF, which is a reasonable value for the setup geometry and wiring. Taking $C_d$= 0 pF (estimated to be a few pF), equation 2 and 3 now give a rise time $T_h$ = 285 ns, which is close to $T_m$. A non-zero value for $C_d$ would only increase $T_h$. Therefore, this analysis strongly indicates that the bandwidth of the HEMT post-amplifier determines the measured rise time. $T_h$ can be reduced by reducing $C_c$ and $R_h$. There is however a trade-off in signal amplitude, because a reduction of $R_h$ also reduces $V_h$.